\documentclass[5p]{elsarticle}
\usepackage{graphicx}
\usepackage{epstopdf}
\usepackage{amssymb}\usepackage{url}
\usepackage{natbib}
\usepackage{url}
\bibpunct{(}{)}{;}{a}{}{,}

\journal{Icarus}

\begin{document}

\begin{frontmatter}


\title{Observations of the 18-cm OH lines of comet 103P/Hartley~2 at
Nan\c{c}ay in support to the \textit{EPOXI} and \textit{Herschel}
missions}

\author[lesia]{Jacques~Crovisier\corref{cor1}}
\ead{jacques.crovisier@obspm.fr}
\cortext[cor1]{Corresponding author.}
\author[lesia]{Pierre~Colom}
\author[lesia]{Nicolas~Biver}
\author[lesia]{Dominique~Bockel\'ee-Morvan} 
\author[bologna,ESO]{J\'er\'emie~Boissier}


\address[lesia]{LESIA -- Observatoire de Paris, CNRS, UPMC,
Universit\'e Paris-Diderot, 5 place Jules Janssen, 92195 Meudon,
France}
\address[bologna]{Istituto di Radioastronomia -- INAF, Via Gobetti
101, Bologna, Italy}
\address[ESO]{ESO, Karl Schwarzchild Str.  2, 85748 Garching bei
M\"{u}nchen, Germany}


\date{Draft \today}



\begin{abstract}

The 18-cm radio lines of the OH radical were observed in comet
103P/Hartley 2 with the Nan\c{c}ay radio telescope in support to its
flyby by the \textit{EPOXI} mission and to observations with the
\textit{Herschel Space Observatory}.  The OH lines were detected from
24 September to 15 December 2010.  These observations are used to
estimate the gas expansion velocity within the coma to $0.83 \pm 0.08$~km~s$^{-1}$ in October 2010.  The water production increased steeply
but progressively before perihelion, and reached $1.9 \pm 0.3 \times
10^{28}$~s$^{-1}$ just before the EPOXI flyby

\end{abstract}

\end{frontmatter}

\section{Introduction}

The Jupiter-family comet 103P/Hartley 2 passed perihelion on 28
October 2010 at $q = 1.059$~AU. It made an exceptional close approach
to the Earth just before perihelion on 21 October at $\Delta =
0.12$~AU, providing us with the best opportunity to observe a
Jupiter-family comet since 73P/Schwassmann-Wachmann 3 in May 2006
\citep{colo+:2006,weaver:2006}.  It was the target of the
\textit{EPOXI} space mission, with a flyby at 700~km on 4.57 (UT)
November 2010 \citep{ahea+:2011}.  For these two reasons, this comet
was the object of an international campaign of observations
\citep{meec+:2011}.  It was also the primary target of the cometary
programme for the \textit{Herschel Space Observatory}
\citep{hart+:2009,hart+:2011_nature}.

The observation of the 18-cm radio lines of the OH radical is a
convenient way to measure the cometary water production rate and its
evolution \citep{desp+:1981, crov+:2002}.  We report here on
supporting observations of these lines in comet 103P/Hartley 2 made
with the Nan\c{c}ay radio telescope.

\begin{figure*}
\centerline{\includegraphics[height=0.86\hsize,angle=270.]{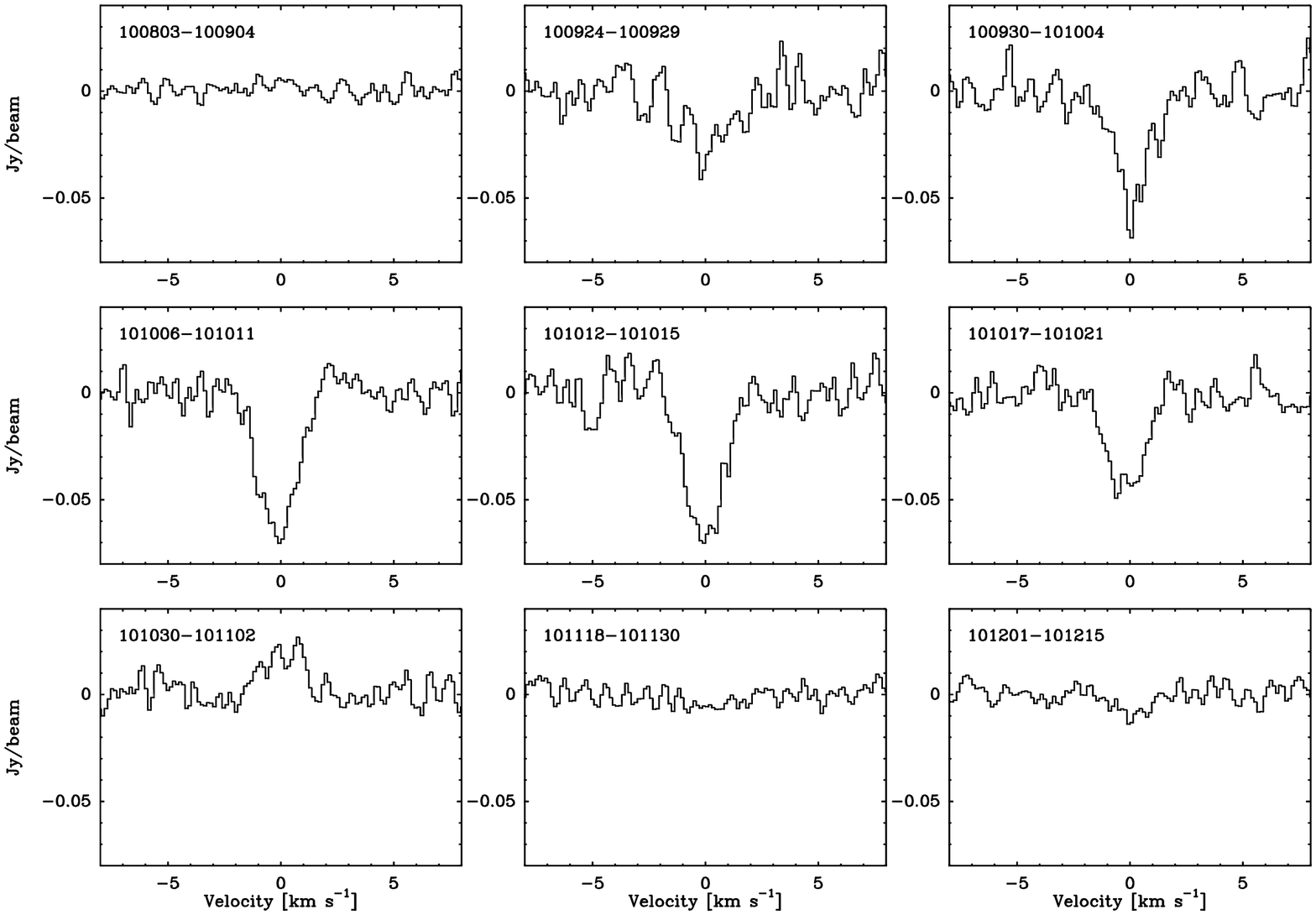}}
\caption{The integrated OH spectra of comet 103P/Hartley 2 (average of
the 1665 and 1667 MHz lines), as listed in Table~\ref{table_103p}
(bottom part).}
  \label{fig:103p}
\end{figure*}

\begin{table*}
\caption{Parameters of the OH spectra of 103P/Hartley 3: individual
spectra recorded between 1st and 21 October, and integrated
spectra for the entire apparition.  For a full log of the spectra, see
\texttt{http://www.lesia.obspm.fr/planeto/cometes/basecom/HT/indexht.html}.}
\label{table_103p}
\centerline{
{\small
\begin{tabular}{lccrrrcrrrrr}
\\
\hline
  date       & $\Delta$ & $r_h$ &$\dot{r}_h$ & $i$ &   $i$ & $T_{bg}$ & $\int S dv$ &           $S$ &             $V_o$ &       $\Delta V$ &       $Q$[OH] \\
 {[yymmdd]}  &  [AU] & [AU] & [km/s] &  (1)   &  (2)   &   [K]    &  [mJy km/s] &         [mJy] &            [km/s] &           [km/s] & [$10^{28}$ s$^{-1}$] \\
\hline
    10/10/ 1.00  &  0.18 & 1.12 &  --7.9 & --0.31 & --0.37 &  3.7 &   --88 $\pm$ 21 & --51 $\pm$ 10 &   0.71 $\pm$ 0.19 &  1.71 $\pm$ 0.48 &  0.7 $\pm$ 0.2 \\
    10/10/ 2.00  &  0.18 & 1.12 &  --7.7 & --0.31 & --0.37 &  3.6 &   --68 $\pm$ 19 & --46 $\pm$  9 & --0.15 $\pm$ 0.20 &  1.64 $\pm$ 0.50 &  0.7 $\pm$ 0.2 \\
    10/10/ 3.01  &  0.17 & 1.11 &  --7.4 & --0.31 & --0.37 &  3.6 &  --112 $\pm$ 23 & --49 $\pm$ 10 & --0.07 $\pm$ 0.21 &  2.09 $\pm$ 0.51 &  0.9 $\pm$ 0.2 \\
    10/10/ 4.01  &  0.17 & 1.11 &  --7.2 & --0.31 & --0.37 &  3.6 &  --134 $\pm$ 18 & --74 $\pm$ 10 &   0.08 $\pm$ 0.11 &  1.48 $\pm$ 0.27 &  0.9 $\pm$ 0.1 \\
    10/10/ 6.03  &  0.16 & 1.10 &  --6.6 & --0.30 & --0.35 &  3.6 &   --94 $\pm$ 20 & --52 $\pm$  9 & --0.06 $\pm$ 0.17 &  2.01 $\pm$ 0.43 &  0.9 $\pm$ 0.2 \\
    10/10/ 7.03  &  0.15 & 1.10 &  --6.4 & --0.29 & --0.34 &  3.6 &  --138 $\pm$ 17 & --62 $\pm$  8 & --0.18 $\pm$ 0.15 &  2.20 $\pm$ 0.37 &  1.1 $\pm$ 0.1 \\
    10/10/ 9.05  &  0.15 & 1.09 &  --5.8 & --0.27 & --0.32 &  3.6 &  --141 $\pm$ 17 & --77 $\pm$  9 & --0.11 $\pm$ 0.11 &  1.77 $\pm$ 0.27 &  1.1 $\pm$ 0.1 \\
    10/10/10.06  &  0.14 & 1.09 &  --5.6 & --0.26 & --0.31 &  3.7 &  --162 $\pm$ 15 & --87 $\pm$  7 & --0.21 $\pm$ 0.08 &  1.86 $\pm$ 0.20 &  1.4 $\pm$ 0.1 \\
    10/10/11.07  &  0.14 & 1.09 &  --5.3 & --0.25 & --0.30 &  3.8 &   --88 $\pm$ 18 & --62 $\pm$ 13 & --0.07 $\pm$ 0.15 &  1.54 $\pm$ 0.35 &  0.8 $\pm$ 0.2 \\
    10/10/12.08  &  0.13 & 1.08 &  --5.0 & --0.24 & --0.29 &  3.9 &  --186 $\pm$ 19 & --75 $\pm$  9 &   0.14 $\pm$ 0.13 &  2.27 $\pm$ 0.34 &  1.6 $\pm$ 0.2 \\
    10/10/13.09  &  0.13 & 1.08 &  --4.7 & --0.23 & --0.27 &  3.8 &  --101 $\pm$ 16 & --68 $\pm$  9 & --0.06 $\pm$ 0.12 &  1.71 $\pm$ 0.29 &  1.2 $\pm$ 0.2 \\
    10/10/14.10  &  0.13 & 1.08 &  --4.4 & --0.22 & --0.26 &  3.8 &  --134 $\pm$ 21 & --82 $\pm$ 10 & --0.16 $\pm$ 0.11 &  1.68 $\pm$ 0.27 &  1.6 $\pm$ 0.2 \\
    10/10/15.10  &  0.13 & 1.07 &  --4.1 & --0.21 & --0.24 &  4.2 &  --123 $\pm$ 18 & --70 $\pm$  8 & --0.04 $\pm$ 0.11 &  1.73 $\pm$ 0.26 &  1.4 $\pm$ 0.2 \\
    10/10/17.12  &  0.12 & 1.07 &  --3.5 & --0.18 & --0.20 &  3.6 &   --99 $\pm$ 17 & --61 $\pm$  9 & --0.11 $\pm$ 0.15 &  1.66 $\pm$ 0.37 &  1.7 $\pm$ 0.3 \\
    10/10/18.13  &  0.12 & 1.07 &  --3.2 & --0.16 & --0.18 &  3.7 &   --79 $\pm$ 15 & --47 $\pm$  8 &   0.03 $\pm$ 0.16 &  1.83 $\pm$ 0.39 &  1.6 $\pm$ 0.3 \\
    10/10/19.13  &  0.12 & 1.07 &  --2.9 & --0.14 & --0.15 &  3.5 &   --64 $\pm$ 20 & --50 $\pm$ 11 & --0.11 $\pm$ 0.14 &  1.09 $\pm$ 0.33 &  1.2 $\pm$ 0.4 \\
    10/10/20.14  &  0.12 & 1.06 &  --2.6 & --0.12 & --0.12 &  3.5 &   --72 $\pm$ 15 & --43 $\pm$  8 & --0.50 $\pm$ 0.14 &  1.61 $\pm$ 0.35 &  1.7 $\pm$ 0.4 \\
    10/10/21.15  &  0.12 & 1.06 &  --2.2 & --0.10 & --0.10 &  3.5 &  --109 $\pm$ 13 & --47 $\pm$  6 & --0.16 $\pm$ 0.14 &  2.04 $\pm$ 0.35 &  2.5 $\pm$ 0.3 \\
\hline
  100803--100904 &  0.53 & 1.42 & --14.1 &   0.25 &   0.17 &  3.3 &      6 $\pm$  4 &               &                   &                  &      $  < 0.3$ \\
  100924--100929 &  0.21 & 1.14 &  --8.8 & --0.27 & --0.32 &  3.5 &   --68 $\pm$  8 & --27 $\pm$  4 &   0.12 $\pm$ 0.19 &  2.62 $\pm$ 0.48 &  0.7 $\pm$ 0.1 \\
  100930--101004 &  0.18 & 1.12 &  --7.5 & --0.31 & --0.37 &  3.6 &  --103 $\pm$ 10 & --53 $\pm$  5 &   0.12 $\pm$ 0.08 &  1.74 $\pm$ 0.21 &  0.8 $\pm$ 0.1 \\
  101006--101011 &  0.15 & 1.09 &  --5.9 & --0.28 & --0.33 &  3.7 &  --131 $\pm$  7 & --70 $\pm$  4 & --0.14 $\pm$ 0.05 &  1.87 $\pm$ 0.13 &  1.1 $\pm$ 0.1 \\
  101012--101015 &  0.13 & 1.08 &  --4.5 & --0.23 & --0.26 &  3.9 &  --132 $\pm$  9 & --72 $\pm$  4 & --0.04 $\pm$ 0.06 &  1.84 $\pm$ 0.15 &  1.4 $\pm$ 0.1 \\
  101017--101021 &  0.12 & 1.07 &  --2.8 & --0.14 & --0.15 &  3.6 &   --86 $\pm$  7 & --46 $\pm$  3 & --0.20 $\pm$ 0.07 &  1.79 $\pm$ 0.18 &  1.8 $\pm$ 0.1 \\
  101030--101102 &  0.14 & 1.06 &    1.1 &   0.04 &   0.10 &  3.4 &     42 $\pm$  7 &   20 $\pm$  3 &   0.16 $\pm$ 0.17 &  1.97 $\pm$ 0.43 &  1.7 $\pm$ 0.3 \\
  101118--101130 &  0.25 & 1.12 &    7.8 & --0.08 & --0.03 &  3.3 &   --14 $\pm$  4 &  --6 $\pm$  2 &   0.08 $\pm$ 0.48 &  3.54 $\pm$ 1.23 &  1.2 $\pm$ 0.4 \\
  101201--101215 &  0.32 & 1.19 &   10.6 & --0.17 & --0.20 &  3.4 &   --22 $\pm$  4 & --14 $\pm$  3 &   0.21 $\pm$ 0.16 &  1.50 $\pm$ 0.40 &  0.6 $\pm$ 0.1 \\
  101217--101230 &  0.41 & 1.30 &   12.8 & --0.14 & --0.21 &  3.4 &    --5 $\pm$  5 &               &                   &                  &     $  <  0.5$ \\
  110103--110131 &  0.60 & 1.51 &   14.8 & --0.08 & --0.10 &  3.4 &      2 $\pm$  3 &               &                   &                  &      $ <  0.9$ \\
\hline
\\
\end{tabular}  }}
{\scriptsize
The display of the Table follows that used by \citet{crov+:2002}.  The
respective columns are the date of the observation (or the date range
for the integrated spectra listed in the bottom part of the table);
the geocentric distance $\Delta$; the heliocentric distance $r_h$; the
heliocentric radial velocity $\dot{r}_h$; the expected inversion $i$
of the OH maser according to (1) \citet{desp+:1981} and (2)
\citet{schl-ahea:1988}; the background brightness temperature
$T_{bg}$; the integrated line area $\int S dv$; the results of a
Gaussian fit to the line: line intensity $S$, line central velocity
$V_o$ and line width at half-maximum $\Delta V$; the OH production
rate $Q$[OH] according to the Haser-equivalent model with collisional
quenching described in \citet{crov+:2002}.  }
  \end{table*}

\section{Observations}

The Nan\c{c}ay radio telescope is a meridian instrument with a $3.5
\times 18$ arcmin field of view at 18-cm wavelength.  The other
characteristics of the telescope and the methodology used for the
cometary observations and their reduction were described by
\citet{crov+:2002}.

\begin{figure}[ht]
\includegraphics[height=\hsize,angle=270.]{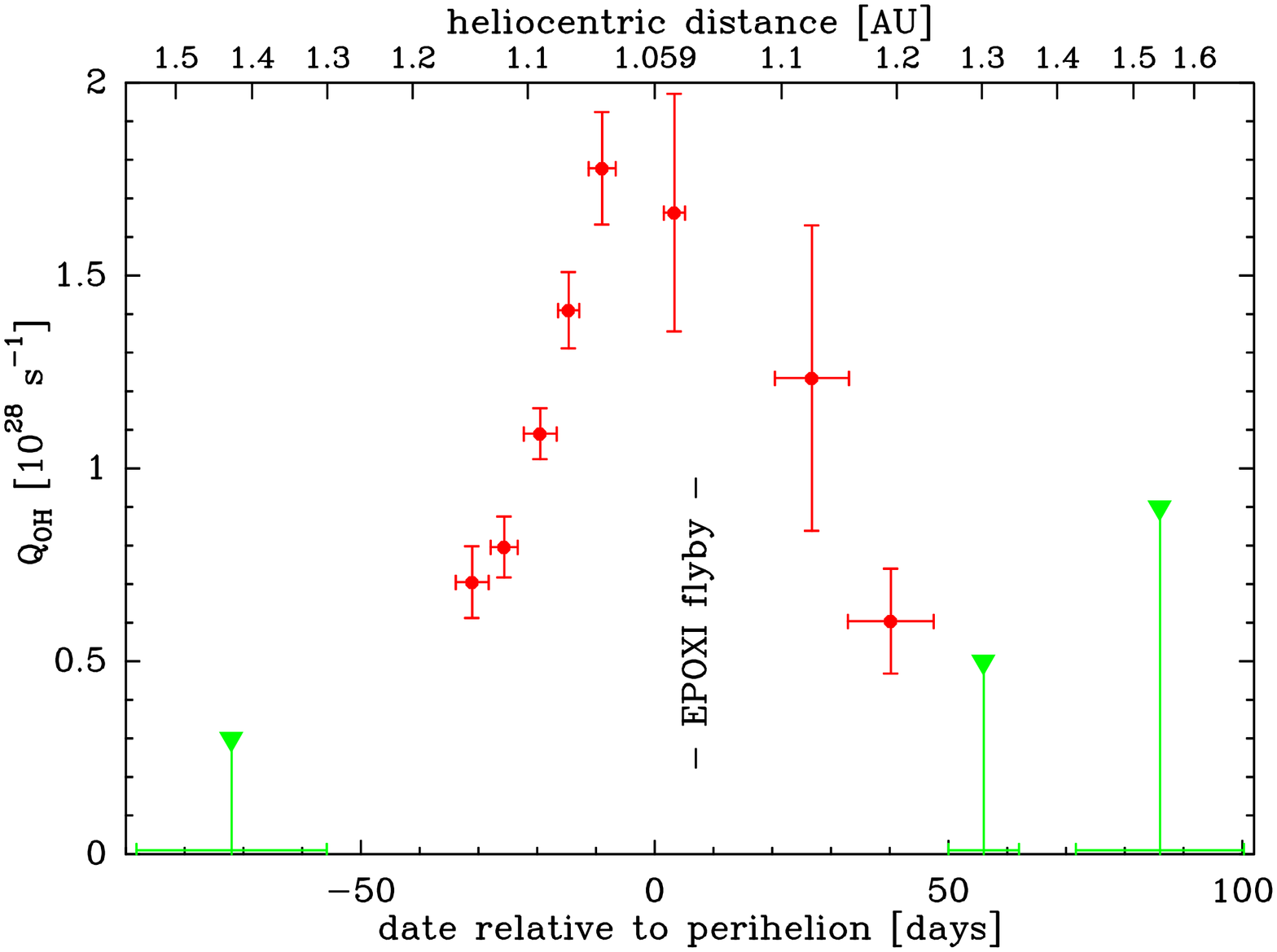}
\caption{Long-term evolution of the OH production rate of 103P/Hartley
2, from averages of Nan\c{c}ay spectra over several days.  Triangles
are upper limits.}
  \label{fig:103p_qoh_long}
\end{figure}

\sloppy Observations of comet 103P/Hartley 2 at Nan\c{c}ay began on 3
August 2010 and were scheduled until the end of January 2011.  The
first detection of 18-cm OH lines occurred at the end of September
with a production rate $Q$[OH] = $0.7 \pm 0.1 \times 10^{28}$~s$^{-1}$
at a heliocentric distance $r_h = 1.14$~AU \citep{crov+:2010-IAUC}.
The comet was detected until 15 December 2010, with episodic gaps due
to telescope scheduling, technical problems, or low OH-maser
inversion.  Excerpts of the results for the 1--21 October period are
listed in the top part of Table~\ref{table_103p}.  Results for
integrated spectra selected over the entire apparition are listed in
the bottom part of Table~\ref{table_103p} and shown in
Fig.~\ref{fig:103p}.  The long-term evolution of the OH production
rate is shown in Fig.~\ref{fig:103p_qoh_long}.  For a full day-by-day
report, see
\texttt{http://www.lesia.obspm.fr/planeto/cometes/ basecom/HT/indexht.html}.

\section{Results and discussion}

The OH production rates reported in Table~\ref{table_103p} and by
\citet{crov+:2010-IAUC} are based on a Haser-equivalent model
\citep{comb+dels:1980}.  The excitation of the OH radical assumes the
$\dot{r}_h$-dependent maser inversion of the model of
\citet{desp+:1981} and takes into account collisional quenching.  An
enhancement of the continuum background as the comet passed across the
galactic plane around 15 October was also taken into account as
explained by \citet{crov+:2002}.

\subsection{Line shape and gas expansion velocity}

Our Haser-equivalent model uses standard values of 0.80~km~s$^{-1}$
for the OH-parent expansion velocity $V_{p}$ and 0.95~km~s$^{-1}$ for
the OH ejection velocity at photodissociation \citep{crov+:2002}.
However, $V_{p}$ may be smaller for this weakly productive comet,
leading to an over-evaluation of the OH production.  Indeed, from
\citet{tsen+:2007} who undertook a statistical study of the OH line
shapes from the Nan\c{c}ay data, we expect $V_{p} \approx
0.71$~km~s$^{-1}$ for a comet with $Q_{p} \approx 10^{28}$~s$^{-1}$ at
$r_{h} \approx 1.09$~AU. The real value may still be smaller since
we are sampling a smaller inner region in this close-by comet.

\begin{figure}[ht]
\centering
\includegraphics[height=0.9\hsize,angle=270.]{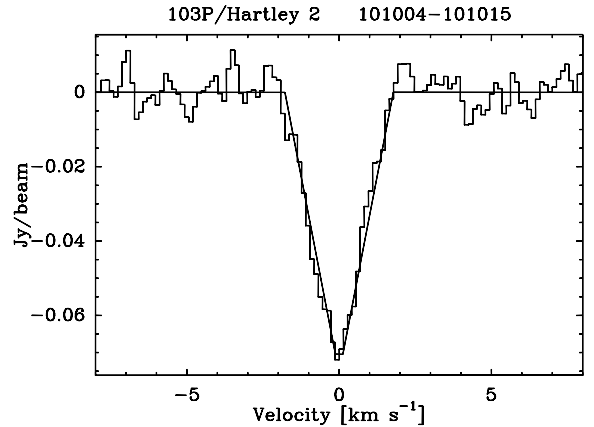}
\caption{A trapezium shape fit to the OH lines of comet 103P/Hartley 2
(average of 4 to 15 October 2010).  An OH-parent expansion velocity of
$0.83 \pm 0.08$ km~s$^{-1}$ was derived using the method of
\citet{bock+:1990}.}
  \label{fig:103p_trap}
\end{figure}

We thus re-evaluated $V_{p}$ following the trapezium method of
\citet{bock+:1990}, from the average spectrum of 4--15 October
obtained at $r_{h} = 1.09$~AU and $\Delta = 0.14$~AU
(Fig.~\ref{fig:103p_trap}) (this method requires a line shape observed
with a high signal-to-noise ratio and can only be applied to this part
of our observations).  We derive $V_{p} = 0.83 \pm 0.08$ km~s$^{-1}$.
This does not make a drastic difference with the expected value from
\citet{tsen+:2007} and agrees well with the expansion velocities
derived from other molecular lines observed at radio wavelengths
\citep[][ and in preparation]{bive+:2011}.

\subsection{The radio OH observations at the time of the \textit{EPOXI} flyby}  

Comet 103P/Hartley~2 could be detected in the period 30 October--2
November, just after perihelion and just before the \textit{EPOXI}
flyby, when $\dot{r}_{h} \approx 1.1$~km~s$^{-1}$, corresponding to a
weakly positive OH inversion (0.04--0.10).  The OH production rate is
then estimated to $1.7 \pm 0.3 \times 10^{28}$~s$^{-1}$.

No observation was possible on 3 and 4 November due to a technical
failure.  At the very time of the \textit{EPOXI} flyby (4 November
2010), the expected OH inversion is small and its predictions differ
between the \citet{desp+:1981} and \citet{schl-ahea:1988} models (0.02
and 0.05, respectively).  The interpretation of the radio OH
observations in terms of production rates is thus difficult at this
moment.

Indeed, the detection of comet 103P/Hartley~2 at Nan\c{c}ay at the end
of October was favoured by the small geocentric distance.  For
comparison, comet 9P/Tempel~1 (the target of the \textit{Deep Impact}
mission) had similar $\dot{r}_{h}$ and maser inversion around 5 July
2005.  It could be then detected with the 100-m Green Bank telescope
thanks to a long integration \citep{howe+:2007}, but not at Nan\c{c}ay
which could only integrate one hour per day \citep{bive+:2007-Icarus}.
With a comparable OH production rate, 9P/Tempel~1 was then at at
$\Delta = 0.90$~AU, whereas 103P/Hartley~2 was much closer at $\Delta
= 0.14$~AU.

\subsection{Comparison with other production rates}

\begin{table*}
\caption{Water production rates measured in comet 103P/Hartley ~2
close to the time of the \textit{EPOXI} flyby.}
\label{table_rates}
\centerline{
\begin{tabular}{llccl}
\\
\hline
Method & instrument & date & $Q$[H$_2$O]$^{a)}$ & reference \\
       &            & 2010 & $10^{28}$ s$^{-1}$ & \\
\hline
H$_2$O sub-mm & \textit{Odin}           & 29 Oct.--1 Nov.  & 0.75--1.24    & \citet{bive+:2010-IAUC,bive+:2011} \\
H$_2$O sub-mm & \textit{Herschel}/HIFI  & 30 Oct.          & $1.0  \pm0.2$ & \citet{lis+:2010-IAUC} \\
H$_2$O sub-mm & \textit{Herschel}/PACS  & 4 Nov.           & 1.2           & \citet{meec+:2011} \\
H$_2$O IR     & Keck/NIRSPEC            & 4 Nov.           & 0.77--1.57    & \citet{dell+:2010,dell+:2011} \\
H$_2$O IR     & Keck/NIRSPEC            & 22 Oct.--16 Nov. & 0.65--0.84    & \citet{mumm+:2010,mumm+:2011} \\
Ly-$\alpha$   & \textit{SOHO}/SWAN      & $\approx$ 4 Nov. & $0.85$        & \citet{comb+:2011} \\
OH near UV    &  Lowell observatory     & 31 Oct.          & 1.15          & \citet{knig-schl:2012} \\
OH 18-cm      & Nan\c{c}ay              & 30 Oct.--2 Nov.  & $1.9\pm 0.3$  & this work \\
\hline
\\
\end{tabular}
}
{\scriptsize $^{a)}$ The listed production rates include both the short-term
temporal variations and the published errors.  The OH production rate
has been multiplied by 1.1 to account for the photolytic branching
ratio.}
\end{table*}

The water production was directly measured in this comet from
submillimetric rotational transitions from space with \textit{Odin}
\citep{bive+:2010-IAUC, bive+:2011} and with the three instruments of
\textit{Herschel} \citep{lis+:2010-IAUC,hart+:2011_nature,meec+:2011},
from ro-vibrational lines in the infrared from the ground
\citep{dell+:2010,dell+:2011,mumm+:2010,mumm+:2011}.  It was
indirectly measured from the observation of its photodissociation
products: of OH in the near-UV \citep{knig-schl:2012}, of H from the
Ly-$\alpha$ line observed by \textit{SOHO}/SWAN \citep{comb+:2011}.

Published water production rates from these concurrent observations,
for dates close to the \textit{EPOXI} flyby, are listed in
Table~\ref{table_rates}.  They range from 0.6 to $1.6 \times 10^{28}$
s$^{-1}$.  The Nan\c{c}ay measurement is on the high side of this
spread.  The importance of the spread may be attributed to the
short-term variation of the comet activity linked to its rotation (see
below), but also to the idiosyncrasies of the various models used to
interpret the different methods of observation. The large Nan\c{c}ay
beam is also more sensitive to water outgassed from icy grains.

\subsection{Short- and long-term variations}

\begin{figure}[ht]
\includegraphics[height=\hsize,angle=270.]{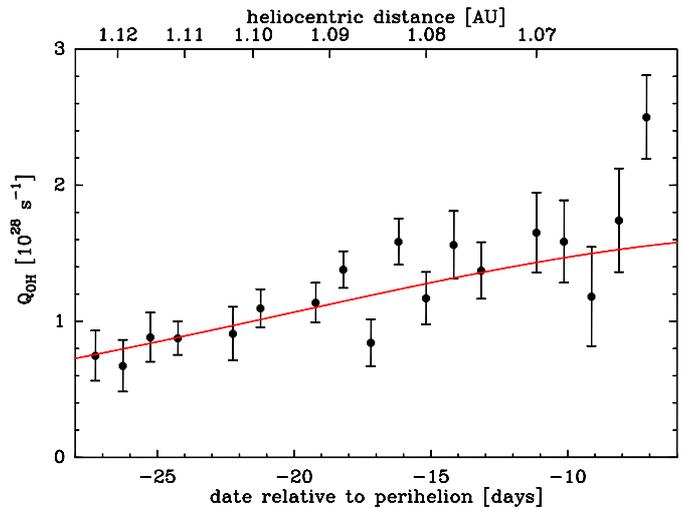}
\caption{Evolution of the OH production rate of 103P/Hartley 2 between
1 and 21 October 2010 (this period has been selected because the
signal on a single day of observation had a significant
signal-to-noise ratio). A power-law model $r_h^{-13}$ is superimposed.}
  \label{fig:103p_qoh_short}
\end{figure}

From many contemporaneous observations --- visible imaging,
spectrophotometry and radar monitoring --- a comet rotation period of
17--18 hours was derived
\citep{jehi+:2010,harm+:2011,knig-schl:2011_AJ,sama+:2011}.  A strong
modulation of the production rates of gas species was observed with
the same period \citep{drah+:2011}.  For water, the modulation had an
amplitude of nearly a factor of two, as was observed by e.g.
\textit{Odin} \citep{bive+:2011}.

The short-term variation of the OH production rate observed at
Nan\c{c}ay in October 2010 is shown in Fig.~\ref{fig:103p_qoh_short}.
One can see a steady increase of the production rate as the comet was
approaching the Sun, but no sign of a periodic modulation.  This is
not unexpected: with a 1-hour observation every day at Nan\c{c}ay, the
time variation of the OH production rate was badly sampled.  Moreover,
temporal variations are averaged over the large beam of Nan\c{c}ay
($3.5 \times 18$~arcmin) and smoothed by the progressive
photodissociation of water which has a lifetime of approximately one
day.  It is thus delusive to try to recover a period of the order of
one day from the Nan\c{c}ay data, although the 7-day period of comet
1P/Halley could be successfully retrieved in the past
\citep{colo-gera:1988}.

In the month preceding perihelion, the water production was steeply
increasing (Figs \ref{fig:103p_qoh_long} and
\ref{fig:103p_qoh_short}), roughly following an $r_h^{-13}$ law.  The
post-perihelion evolution shows a similarly steep decrease (but not so
well constrained due to larger errors).  These variations can
difficultly be explained by the small variation in the heliocentric
distance.  A seasonal effect may be suspected, which could be further
investigated when the rotational state of the comet and especially its
pole orientation will be precisely known.

The water production rate preceding perihelion measured by
\textit{SOHO}/SWAN also shows an increase by a similar amount.  But
this increase appears to be almost entirely due to a sudden jump ---
by a factor $\approx 2.5$ --- 27 days before perihelion
\citep{comb+:2011}.  This jump is not present in the Nan\c{c}ay data
for which the rise is progressive.

\section{Conclusion}

\begin{itemize}

\item The 18-cm lines of OH were monitored in comet 103P/Hartley 2
with the Nan\c{c}ay radio telescope from August 2010 to January 2011.
They were detected from 24 September to 15 December 2010.

\item From the line shapes, we derive an expansion velocity $V_{p} =
0.83 \pm 0.08$ km~s$^{-1}$ for the gas coma in October.  This is comparable to
what is observed for comets with comparable gas production rates at
similar heliocentric distances.

\item The water production rate is estimated to $1.9\pm 0.3\times
10^{28}$~s$^{-1}$ just before the \textit{EPOXI} flyby.

\item The timing of our meridian observations was not appropriate to
retrieve a possible modulation of the comet activity by its rotation if the period is as short as $\approx$ one day.
The water production was observed to increase steeply, but
progressively, in the month preceding perihelion.

\end{itemize}

\section*{Acknowledgements} 

\sloppy The Nan\c{c}ay Radio Observatory is the Unit\'e scientifique
de Nan\c{c}ay of the Observatoire de Paris, associated as Unit\'e de
service et de recherche (USR) No B704 to the French Centre national de
la recherche scientifique (CNRS).  Its upgrade was financed jointly by
the Conseil r\'egional of the R\'egion Centre in France, the CNRS and
the Observatoire de Paris.  Part of the research leading to these
results received funding from the European Community's Seventh
Framework Programme (FP7/2007--2013) under grant agreement No.
229517.

\bibliographystyle{icarus}
\bibliography{103P_nancay_icarus_v2}

\end{document}